\documentclass[aps,tightlines]{revtex4}

\usepackage{amsfonts}
\usepackage{amsmath}
\usepackage{amssymb}
\usepackage{graphicx}

\begin{document}

\title[Partial $K-$way negativities...]{Partial $K-$way Negativities of Pure
Four qubit Entangled States}
\author{S. Shelly Sharma}
\email{shelly@uel.br}
\affiliation{Depto. de F\'{\i}sica, Universidade Estadual de Londrina, Londrina
86051-990, PR Brazil }
\author{N. K. Sharma}
\email{nsharma@uel.br}
\affiliation{Depto. de Matem\'{a}tica, Universidade Estadual de Londrina, Londrina
86051-990 PR, Brazil }
\thanks{}

\begin{abstract}
It has been shown by Versraete et. al [F. Versraete, J. Dehaene, B. De Moor,
and H. Verschelde, Phys. Rev. A65, 052112 (2002)] that by Stochastic local
operations and classical communication (SLOCC), a pure state of four qubits
can be transformed to a state belonging to one of a set of nine families of
states. By using selective partial transposition, we construct partial $K-$%
way negativities to measure the genuine $4-$partite, tripartite, and
bi-partite entanglement of single copy states belonging to the nine families
of four qubit states. Partial $K-$way negativities are polynomial functions
of local invariants characterizing each family of states, as such,
entanglement monotones.
\end{abstract}

\maketitle

Detection and measurement of multipartite entanglement is an important open
question. Different parts of an N-partite composite system may be entangled
to each other in distinctly different ways. In particular, four qubit ($ABCD$%
) states may have $4-$partite, tri-partite, and bi-partite entanglement. The
bipartite entanglement, may in turn be present for a given pair of qubits or
for more than one qubit pairs. Similarly, possible candidates for tripartite
entanglement are subsystems, $ABC$, $ABD$, $ACD$, and $BCD.$ A four qubit
state may have four-partite entanglement generated by three qubit coherence
or tripartite entanglement due to two qubit quantum correlations. Negativity 
\cite{zycz98} based on Peres Horodecki PPT separability criterion \cite%
{pere96,horo96} has been shown to be an entanglement monotone \cite%
{vida02,eise01}. We proposed a characterization of three qubit states \cite%
{shar08,shar082} based on global negativities and partial $K-$way
negativities ($2\leq K\leq 3$). The $K-$way partial transpose with respect
to a subsystem of an N-partite composite system is constructed by partial
transposition subject to specific constraints on transposed matrix elements
for each value of $K$ $\left( 2\leq K\leq N\right) $. The $K-$way partial
transpose may also be constructed for a given set of $K$ subsystems. The $K-$%
way negativity ($2\leq K\leq N)$, defined as the negativity of $K-$way
partial transpose, quantifies the $K-$way coherences of the composite
system. The underlying idea of selective transposition to construct a $K-$%
way partial transpose with respect to a subsystem, presented for the first
time in ref. \cite{shel06} and then applied in ref. \cite{shar07}, shifts
the focus from $K-$subsystems to $K-$way coherences of the composite system.
By $K-$way coherences, we mean the quantum correlations responsible for GHZ
state like entanglement of a K-partite system. For an N-partite entangled
state, the partial K-way negativity is the contribution of a $K-$way partial
transposes ($2\leq K\leq N$) to the global negativity.

In this article, we present analytical expressions and numerical
calculations of partial $K-$way negativities for single copy pure four qubit
states. Entanglement being a nonlocal property of the composite system, it
cannot be increased by local operations and classical communication (LOCC).
Versraete et al. \cite{vers02} have shown that a pure state of four qubits
can be transformed, by Stochastic local operations and classical
communication (SLOCC), to a state belonging to one of a set of nine families
of states, corresponding to nine distinct ways of entangling four qubits. We
use the global and $K-$way partial transposes, to construct measures of
genuine $4-$partite, tri-partite, and bi-partite entanglement for all the
nine families of states. It has been pointed out by Versraete et al. \cite%
{vers02}, that the polynomial functions of local invariants characterizing
each family are entanglement monotones. The calculated global and partial $%
K- $way negativities are polynomial functions of local invariants for each
family of states, as such entanglement monotones. The nine families of
states are grouped in to two classes, with (i) class I containing states for
which partial three-way negativity is zero and (ii) class II states
characterised by nonzero partial three-way negativity.

The necessary definitions to construct global and $K-$way partial transpose
are given in section I, and partial K-way negativities defined in section
II. Calculation of pairwise and three qubit entanglement for specific groups
of qubits is discussed in section III. Comments on notation and
classification of nine families of states are included in section IV. In
section V, entanglement of a general four qubit state is analyzed and the
entanglement measures of states belonging to class I are shown to be related
to those for the general four qubit state. Monogamy inequalities for the
states in Class I are also presented. Section VI presents the states with
non zero partial three-way negativities. The results are summarized in
section VII.

\section{The Global and K-way partial transpose of 4-qubit state}

We consider qubits one, two, three and four located at labs $A$, $B$, $C$
and $D$, respectively, constituting a composite system $ABCD$ in state $%
\widehat{\rho }$ . The Hilbert space, $C^{2^{4}}$, associated with the
quantum system is spanned by basis vectors of the form $\left\vert
i_{1}i_{2}i_{3}i_{4}\right\rangle ,$ where $i_{m}\ =0$ or $1,$ for $m=1$ to $%
4$. To simplify the notation we denote the vector $\left\vert
i_{1}i_{2}i_{3}i_{4}\right\rangle $ by $\left\vert
\prod\limits_{m=1}^{4}i_{m}\right\rangle $and write a general four qubit
pure state as 
\begin{equation}
\widehat{\rho }=\sum_{\substack{ i_{1}-i_{4},  \\ j_{1}-j_{4}}}\left\langle
\prod\limits_{m=1}^{4}i_{m}\right\vert \widehat{\rho }\left\vert
\prod\limits_{m=1}^{4}j_{m}\right\rangle \left\vert
\prod\limits_{m=1}^{4}i_{m}\right\rangle \left\langle
\prod\limits_{m=1}^{4}j_{m}\right\vert .  \label{1}
\end{equation}%
The global partial transpose of $\widehat{\rho }$ with respect to qubit $p$
is defined as%
\begin{equation}
\widehat{\rho }_{G}^{T_{p}}=\sum_{\substack{ i_{1}-i_{4},  \\ j_{1}-j_{4}}}%
\left\langle j_{p}\prod\limits_{m=1,m\neq p}^{4}i_{m}\right\vert \widehat{%
\rho }\left\vert i_{p}\prod\limits_{m=1,m\neq p}^{4}j_{m}\right\rangle
\left\vert \prod\limits_{m=1}^{4}i_{m}\right\rangle \left\langle
\prod\limits_{m=1}^{4}j_{m}\right\vert .  \label{ptg}
\end{equation}%
The partial transpose $\widehat{\rho }_{G}^{T_{p}}$ of a state having free
entanglement is non positive. The Global negativity defined as 
\begin{equation}
N_{G}^{p}=\frac{1}{d_{p}-1}\left( \left\Vert \rho _{G}^{T_{p}}\right\Vert
_{1}-1\right) ,
\end{equation}%
measures the entanglement of subsystem $p$ with it's complement. Here $%
\left\Vert \widehat{\rho }\right\Vert _{1}$ is the trace norm of $\widehat{%
\rho }$. The factor $1.0/\left( d_{p}-1\right) $ ensures that the maximum
value of negativity is one. Global negativity vanishes on PPT states.

We label a given matrix element $\left\langle
\prod\limits_{m=1}^{4}i_{m}\right\vert \widehat{\rho }\left\vert
\prod\limits_{m=1}^{4}j_{m}\right\rangle $ by a number $K=\sum%
\limits_{m=1}^{4}(1-\delta _{i_{m},j_{m}}),$ where $\delta _{i_{m},j_{m}}=1$
for $i_{m}=j_{m}$, and $\delta _{i_{m},j_{m}}=0$ for $i_{m}\neq j_{m}$. The $%
K-$way partial transpose ($K>2$) of\ $\rho $ with respect to subsystem $p$
is obtained by selective transposition such that 
\begin{eqnarray}
\left\langle \prod\limits_{m=1}^{4}i_{m}\right\vert \widehat{\rho }%
_{K}^{T_{p}}\left\vert \prod\limits_{m=1}^{4}j_{m}\right\rangle 
&=&\left\langle j_{p}\prod\limits_{m=1,m\neq p}^{4}i_{m}\right\vert \widehat{%
\rho }\left\vert i_{p}\prod\limits_{m=1,m\neq p}^{4}j_{m}\right\rangle , 
\notag \\
\quad \text{if}\quad \sum\limits_{m=1}^{4}(1-\delta _{i_{m},j_{m}})
&=&K,\quad \text{and }\quad \delta _{i_{p},j_{p}}=0  \label{ptk1}
\end{eqnarray}%
and%
\begin{eqnarray}
\left\langle \prod\limits_{m=1}^{4}i_{m}\right\vert \widehat{\rho }%
_{K}^{T_{p}}\left\vert \prod\limits_{m=1}^{4}j_{m}\right\rangle 
&=&\left\langle \prod\limits_{m=1}^{4}i_{m}\right\vert \widehat{\rho }%
\left\vert \prod\limits_{m=1}^{4}j_{m}\right\rangle ,  \notag \\
\quad \text{if}\quad \sum\limits_{m=1}^{4}(1-\delta _{i_{m},j_{m}}) &\neq &K.
\label{ptk2}
\end{eqnarray}%
while%
\begin{eqnarray}
\left\langle \prod\limits_{m=1}^{4}i_{m}\right\vert \widehat{\rho }%
_{2}^{T_{p}}\left\vert \prod\limits_{m=1}^{4}j_{m}\right\rangle 
&=&\left\langle j_{p}\prod\limits_{m=1,m\neq p}^{4}i_{m}\right\vert \widehat{%
\rho }\left\vert i_{p}\prod\limits_{m=1,m\neq p}^{4}j_{m}\right\rangle , 
\notag \\
\quad \text{if}\quad \sum\limits_{m=1}^{4}(1-\delta _{i_{m},j_{m}}) &=&1%
\text{ or }2,\quad \text{and }\quad \delta _{i_{p},j_{p}}=0  \label{pt21}
\end{eqnarray}%
and%
\begin{eqnarray}
\left\langle \prod\limits_{m=1}^{4}i_{m}\right\vert \widehat{\rho }%
_{2}^{T_{p}}\left\vert \prod\limits_{m=1}^{4}j_{m}\right\rangle 
&=&\left\langle \prod\limits_{m=1}^{4}i_{m}\right\vert \widehat{\rho }%
\left\vert \prod\limits_{m=1}^{4}j_{m}\right\rangle ,  \notag \\
\quad \text{if}\quad \sum\limits_{m=1}^{4}(1-\delta _{i_{m},j_{m}}) &\neq &1%
\text{ or }2.  \label{pt22}
\end{eqnarray}%
Recalling that a single qubit matrix element of spin flip operator $\widehat{%
\sigma }_{x}^{m}$ is given by $\left\langle i_{m}\right\vert \widehat{\sigma 
}_{x}^{m}\left\vert j_{m}\right\rangle =(1-\delta _{i_{m},j_{m}})$, $%
\widehat{\rho }_{K}^{T_{p}}$ is obtained by restricting the transposition to
the group of matrix elements of $\widehat{\rho }$ in which the number of
spins being flipped is $K$. In case matrix $\rho $ is a real matrix, the
definition of $\widehat{\rho }_{2}^{T_{p}}$ used here gives the same results
as that used in refs.[\cite{shar07, shar08}]. However, if state operator is
represented by a complex Hermitian matrix, the contribution of $\widehat{%
\rho }_{2}^{T_{p}}$, as defined by Eqs. (\ref{pt21}) and (\ref{pt22}), to
global negativity is different. The correction ensures that, $\widehat{\rho }%
_{G}^{T_{p}}=\widehat{\rho }_{2}^{T_{p}},$ in case the state $\widehat{\rho }
$ has only pairwise entanglement. The $K-$way negativity calculated from $K-$%
way partial transpose of matrix $\rho $ with respect to subsystem $p$, is
defined as $N_{K}^{p}=\frac{1}{d_{p}-1}\left( \left\Vert \rho
_{K}^{T_{p}}\right\Vert _{1}-1\right) $. Using the definition of trace norm
and the fact that $tr(\rho _{K}^{T_{p}})=1$, we get $N_{K}^{p}=\frac{2}{%
d_{p}-1}\sum_{i}\left\vert \lambda _{i}^{K-}\right\vert $, $\lambda _{i}^{K-}
$ being the negative eigenvalues of matrix $\rho _{K}^{T_{p}}$. The
negativity $N_{K}^{p}$ ($K=2$ to $4$), depends on $K-$way coherences and is
a measure of all possible types of entanglement attributed to $K-$ way
coherences. By $K-$ way coherences, we mean the quantum correlations
responsible for genuine $K-$partite entanglement of the composite system.

\section{Partial K-way Negativities}

Global negativity with respect to a subsystem $p$ can be written as a sum of
partial $K-$way negativities. Using $Tr\left( \widehat{\rho }%
_{G}^{T_{p}}\right) =1,$ the negativity of $\widehat{\rho }_{G}^{T_{p}}$ is
given by 
\begin{equation}
{N}_{G}^{p}=-\frac{2}{d_{p}-1}\sum\limits_{i}\left\langle \Psi
_{i}^{G-}\right\vert \widehat{\rho }_{G}^{T_{p}}\left\vert \Psi
_{i}^{G-}\right\rangle =-\frac{2}{d_{p}-1}\sum\limits_{i}\lambda _{i}^{G-}%
\text{,}  \label{2n}
\end{equation}%
where $\lambda _{i}^{G-}$and $\left\vert \Psi _{i}^{G-}\right\rangle $ are,
respectively, the negative eigenvalues and eigenvectors of $\widehat{\rho }%
_{G}^{T_{p}}$. It is straight forward to verify that 
\begin{equation}
\widehat{\rho }_{G}^{T_{p}}=\sum\limits_{K=2}^{4}\widehat{\rho }%
_{K}^{T_{p}}-2\widehat{\rho }.  \label{3n}
\end{equation}%
For a pure state with qubit $p$ separable, the expansion of Eq. (\ref{3n}),
leads to the equality

\begin{eqnarray}
\left\Vert \rho _{G}^{T_{p}}\right\Vert _{1} &=&\sum\limits_{K=2}^{4}\left(
\sum\limits_{i}\left\vert \left\langle \Psi _{i}^{G}\right\vert \widehat{%
\rho }_{K}^{T_{p}}\left\vert \Psi _{i}^{G}\right\rangle \right\vert \right)
-2\sum\limits_{i}\left\vert \left\langle \Psi _{i}^{G}\right\vert \widehat{%
\rho }\left\vert \Psi _{i}^{G}\right\rangle \right\vert  \notag \\
&=&\sum\limits_{K=2}^{4}tr\left( \widehat{\rho }_{K}^{T_{p}}\right)
-2tr\left( \widehat{\rho }\right) =1,
\end{eqnarray}%
where $\Psi _{i}^{G}$ are eigen functions of a positive global partial
transpose $\rho _{G}^{T_{p}}$. On the other hand, when qubit $p$ is
entangled that is ${N}_{G}^{p}>0$, by substituting Eq. (\ref{3n}) in Eq. (%
\ref{2n}), we get%
\begin{equation}
-2\sum\limits_{i}\lambda
_{i}^{G-}=-2\sum\limits_{K=1}^{4}\sum\limits_{i}\left\langle \Psi
_{i}^{G-}\right\vert \widehat{\rho }_{K}^{T_{p}}\left\vert \Psi
_{i}^{G-}\right\rangle +4\sum\limits_{i}\left\langle \Psi
_{i}^{G-}\right\vert \widehat{\rho }\left\vert \Psi _{i}^{G-}\right\rangle .
\end{equation}%
Defining partial $K-$way negativity $E_{K}^{p}$ ($K=2$ to $4$) as 
\begin{eqnarray}
E_{K}^{p} &=&-\frac{2}{d_{p}-1}\sum\limits_{i}\left\langle \Psi
_{i}^{G-}\right\vert \widehat{\rho }_{K}^{T_{p}}\left\vert \Psi
_{i}^{G-}\right\rangle ),  \label{4n} \\
E_{0}^{p} &=&-\frac{2}{d_{p}-1}\sum\limits_{i}\left\langle \Psi
_{i}^{G-}\right\vert \widehat{\rho }\left\vert \Psi _{i}^{G-}\right\rangle ,
\end{eqnarray}%
we may split the global negativity for qubit $p$ as%
\begin{equation}
N_{G}^{p}=\sum\limits_{K=2}^{4}\left( E_{K}^{p}-E_{0}^{p}\right) +E_{0}^{p}.
\label{5n}
\end{equation}%
The eigen vectors of global partial transpose and $K-$way partial transpose
are not the same except when the global partial transpose is equal to $K-$%
way partial transpose. Given that $\widehat{\rho }_{K}^{T_{p}}\left\vert
\Psi _{\mu }^{K}\right\rangle =\beta _{K}^{\mu }\left\vert \Psi _{\mu
}^{K}\right\rangle $, and representing by $\beta _{K}^{\mu -}$ and $\beta
_{K}^{\mu +}$, the negative and positive eigenvalues of $\widehat{\rho }%
_{K}^{T_{p}}$, we may rewrite the partial $K-$way negativity as%
\begin{eqnarray}
E_{K}^{p} &=&\frac{2}{d_{p}-1}\left( \sum_{\mu -}\sum\limits_{i}\left\vert
\beta _{K}^{\mu -}\right\vert \left\vert \left\langle \Psi _{i}^{G-}\right.
\left\vert \Psi _{\mu }^{K-}\right\rangle \right\vert ^{2}\right.  \notag \\
&&\left. -\sum_{\mu +}\sum\limits_{i}\beta _{K}^{\mu +}\left\vert
\left\langle \Psi _{i}^{G-}\right. \left\vert \Psi _{\mu }^{K+}\right\rangle
\right\vert ^{2}\right) .  \label{epk}
\end{eqnarray}%
It follows from Eq. (\ref{epk}) that $E_{K}^{p}>0$, if and only if one or
more eigenvalues of $\widehat{\rho }_{K}^{T_{p}}$ are negative and eigen
functions of $\widehat{\rho }_{K}^{T_{p}}$ have finite overlap with
eigenfunctions of $\widehat{\rho }_{G}^{T_{p}}$ corresponding to negative
eigenvalues of $\widehat{\rho }_{G}^{T_{p}}$. We notice that, in the
limiting case, with all the matrix elements having $\sum\limits_{m=1}^{4}(1-%
\delta _{i_{m},j_{m}})=K,$ and $\delta _{i_{p},j_{p}}=0$ satisfying 
\begin{equation}
\left\langle \prod\limits_{m=1}^{4}i_{m}\right\vert \widehat{\rho }%
\left\vert \prod\limits_{m=1}^{4}j_{m}\right\rangle =\left\langle
j_{p}\prod\limits_{m=1,m\neq p}^{4}i_{m}\right\vert \widehat{\rho }%
\left\vert i_{p}\prod\limits_{m=1,m\neq p}^{4}j_{m}\right\rangle ,
\end{equation}%
we obtain $\widehat{\rho }_{K}^{T_{p}}=\widehat{\rho }$ and $K-$way
coherences, if present cannot be detected by $K-$way partial transpose. Only
positive quantities $\left( E_{K}^{p}-E_{0}^{p}\right) $ ($K>1$) are used to
measure the $K-$way coherences of the system. A positive $K-$way partial
transpose of a pure state represents another pure state $\left\vert \Psi
^{\prime }\right\rangle $ of the system, having larger overlap with a given $%
\left\vert \Psi _{i}^{G-}\right\rangle $ than $\left\vert \Psi \right\rangle 
$, leading to $E_{K}^{p}-E_{0}^{p}\leq 0$. This result follows from the
observation that for a pure separable state $\left\vert \Phi ^{A}\otimes
\Phi ^{B}\right\rangle $, with partial transpose 
\begin{equation*}
\widehat{\rho }_{G}^{T_{A}}=\left\vert \left( \Phi ^{A}\right) ^{\ast
}\otimes \Phi ^{B}\right\rangle \left\langle \left( \Phi ^{A}\right) ^{\ast
}\otimes \Phi ^{B}\right\vert ,
\end{equation*}%
the overlaps satisfy 
\begin{equation}
\left\vert \left\langle \left( \Phi ^{A}\right) ^{\ast }\otimes \Phi
^{B}\right. \left\vert \Phi ^{A}\otimes \Phi ^{B}\right\rangle \right\vert
^{2}-\left\vert \left\langle \left( \Phi ^{A}\right) ^{\ast }\otimes \Phi
^{B}\right. \left\vert \left( \Phi ^{A}\right) ^{\ast }\otimes \Phi
^{B}\right\rangle \right\vert ^{2}\leq 0\text{,}
\end{equation}%
where equality holds, only, when all probability amplitudes in the expansion
of $\Phi ^{A}$ are real. For two and three qubit states in canonical form $%
E_{0}^{p}=0$, as such, the partial negativity $E_{K}^{p}$ ($K=2,3$) turns
out to be the measure of $K-$way entanglement. The necessary condition for a 
$4-$qubit pure state not to have genuine $4-$partite entanglement is that at
least one of the global negativities $N_{G}^{p}$ is zero, where $p$ is one
of the subsystems or one part of a bipartite split of the composite system.

\section{How is the pairwise and three qubit entanglement distributed in a
four qubit state?}

It is common practice to trace out subsystem $AD$ to obtain the entanglement
of pair $BC$. State reduction is an irreversible local operation and it is
believed that the entanglement of the pair $BC$ in the reduced system is
either the same or less than that in the composite system $\widehat{\rho }$.
One can, however, obtain a measure of $2-$way coherences involving a given
pair of subsystems by using a $2-$way partial transpose constructed from the
state operator $\widehat{\rho }$ by restricting the transposed matrix
elements to those for which the state of the third and fourth qubit does not
change. For instance, $\widehat{\rho }_{2}^{T_{A-AB}}$ is obtained from the
matrix $\rho $ by applying the condition%
\begin{eqnarray}
\left\langle i_{1}i_{2}i_{3}i_{4}\right\vert \widehat{\rho }%
_{2}^{T_{A-AB}}\left\vert j_{1}j_{2}i_{3}i_{4}\right\rangle &=&\left\langle
j_{1}i_{2}i_{3}i_{4}\right\vert \widehat{\rho }\left\vert
i_{1}j_{2}i_{3}i_{4}\right\rangle ;\quad  \notag \\
\text{if\quad }\sum\limits_{m=1}^{2}\left( 1-\delta _{i_{m},j_{m}}\right)
&=&1\text{ or }2,\text{ and }\sum\limits_{m=3}^{4}\left( 1-\delta
_{i_{m},j_{m}}\right) =0,  \label{10}
\end{eqnarray}%
and for all other matrix elements%
\begin{equation}
\left\langle i_{1}i_{2}i_{3}i_{4}\right\vert \widehat{\rho }%
_{2}^{T_{A-AB}}\left\vert j_{1}j_{2}j_{3}i_{4}\right\rangle =\left\langle
i_{1}i_{2}i_{3}i_{4}\right\vert \widehat{\rho }\left\vert
j_{1}j_{2}j_{3}i_{4}\right\rangle .
\end{equation}%
The negativity $N_{2}^{A-AB}=\frac{1}{d_{A}-1}\left( \left\Vert \widehat{%
\rho }_{2}^{T_{A-AB}}\right\Vert _{1}-1\right) $ measures the $2-$way
coherences involving the pair of subsystems $AB$. It is easily shown that
for a four qubit system 
\begin{equation}
\widehat{\rho }_{2}^{T_{A}}=\widehat{\rho }_{2}^{T_{A-AB}}+\widehat{\rho }%
_{2}^{T_{A-AC}}+\widehat{\rho }_{2}^{T_{A-AD}}-2\widehat{\rho }.
\label{ro2tsum}
\end{equation}%
Substituting Eq. (\ref{ro2tsum}) in the definition of $E_{2}^{A}$ that is 
\begin{equation}
E_{2}^{A}=-2\sum\limits_{i}\left\langle \Psi _{i}^{G-}\right\vert (\widehat{%
\rho }_{2}^{T_{A-AB}}+\widehat{\rho }_{2}^{T_{A-AC}}+\widehat{\rho }%
_{2}^{T_{A-AD}}-2\widehat{\rho })\left\vert \Psi _{i}^{G-}\right\rangle ,
\end{equation}%
we get the relation%
\begin{equation}
E_{2}^{A}-E_{0}^{A}=\left( E_{2}^{A-AB}-E_{0}^{A}\right) +\left(
E_{2}^{A-AC}-E\right) _{0}^{A}+\left( E_{2}^{A-AD}-E_{0}^{A}\right) ,
\end{equation}%
where $E_{2}^{A-AB}$, $E_{2}^{A-AC}$and $E_{2}^{A-AD}$ are contributions of $%
\widehat{\rho }_{2}^{T_{A-AB}}$, $\widehat{\rho }_{2}^{T_{A-AC}}$, and $%
\widehat{\rho }_{2}^{T_{A-AD}}$ to $E_{2}^{A}$.

We can also construct the partially transposed matrices $\widehat{\rho }%
_{3}^{T_{A-ABC}}$, $\widehat{\rho }_{3}^{T_{A-ABD}}$, and $\widehat{\rho }%
_{3}^{T_{A-ACD}}$ such that 
\begin{equation}
\widehat{\rho }_{3}^{T_{A}}=\widehat{\rho }_{3}^{T_{A-ABC}}+\widehat{\rho }%
_{3}^{T_{A-ABD}}+\widehat{\rho }_{3}^{T_{A-ACD}}-2\widehat{\rho }.
\label{ro3tsum}
\end{equation}%
Here $\widehat{\rho }_{3}^{T_{A-ABC}}$is constructed subject to the
conditions 
\begin{eqnarray}
\left\langle i_{1}i_{2}i_{3}i_{4}\right\vert \widehat{\rho }%
_{3}^{T_{A-ABC}}\left\vert j_{1}j_{2}i_{3}i_{4}\right\rangle &=&\left\langle
j_{1}i_{2}i_{3}i_{4}\right\vert \widehat{\rho }\left\vert
i_{1}j_{2}j_{3}i_{4}\right\rangle ;\quad if\quad \sum\limits_{m=1}^{3}\left(
1-\delta _{i_{m},j_{m}}\right) =3,  \notag \\
\left\langle i_{1}i_{2}i_{3}i_{4}\right\vert \widehat{\rho }%
_{3}^{T_{A-ABC}}\left\vert j_{1}j_{2}j_{3}j_{4}\right\rangle &=&\left\langle
i_{1}i_{2}i_{3}i_{4}\right\vert \widehat{\rho }\left\vert
j_{1}j_{2}j_{3}j_{4}\right\rangle ;\quad \text{for all other matrix elements}%
.
\end{eqnarray}%
Analogous restrictions are applied to construct $\widehat{\rho }%
_{3}^{T_{A-ABD}}$, $\widehat{\rho }_{3}^{T_{A-ACD}}$ and $\widehat{\rho }%
_{3}^{T_{B-BCD}}$ etc. Using Eqs. (\ref{4n}) and (\ref{ro3tsum}), we obtain

\begin{eqnarray}
E_{3}^{A}-E_{0}^{A} &=&-2\sum\limits_{i}\left\langle \Psi
_{i}^{G-}\right\vert (\widehat{\rho }_{3}^{T_{A-ABC}}+\widehat{\rho }%
_{3}^{T_{A-ABD}}+\widehat{\rho }_{3}^{T_{A-ACD}}-3\widehat{\rho })\left\vert
\Psi _{i}^{G-}\right\rangle \\
&=&\left( E_{3}^{A-ABC}-E_{0}^{A}\right) +\left(
E_{3}^{A-ABD}-E_{0}^{A}\right) +\left( E_{3}^{A-ACD}-E_{0}^{A}\right) ,
\end{eqnarray}%
where%
\begin{eqnarray}
E_{3}^{A-ABC} &=&-2\sum\limits_{i}\left\langle \Psi _{i}^{G-}\right\vert (%
\widehat{\rho }_{3}^{T_{A-ABC}}\left\vert \Psi _{i}^{G-}\right\rangle , 
\notag \\
\quad E_{3}^{A-ABD} &=&-2\sum\limits_{i}\left\langle \Psi
_{i}^{G-}\right\vert \widehat{\rho }_{3}^{T_{A-ABD}}\left\vert \Psi
_{i}^{G-}\right\rangle , \\
E_{3}^{A-ACD} &=&-2\sum\limits_{i}\left\langle \Psi _{i}^{G-}\right\vert 
\widehat{\rho }_{3}^{T_{A-ACD}}\left\vert \Psi _{i}^{G-}\right\rangle .\quad
\end{eqnarray}

\section{Entanglement of Four qubits canonical States}

A composite system state $\widehat{\rho }^{\prime },$ obtained from an
N-partite state $\widehat{\rho }$ through local unitary operations, differs
from the former in being characterized by a different set of partial $K-$way
negativities. A canonical state $\widehat{\rho }_{c}$ obtained from $%
\widehat{\rho }$ through entanglement conserving local unitary operations is
a state written in terms of the minimum number of local basis product states 
\cite{cart99}. The coeficients in a canonical form are local invariants. The
partial $K-$way negativities $E_{K}^{p}$ (defined in Eq. (\ref{4n})), when
calculated for a canonical state are functions of local invariants having
unique values, as such qualify to be entanglement monotones for all the
states lying on the orbit. For a canonical state $\widehat{\rho }_{c}$, $%
E_{K}^{p}-E_{0}^{p}$, measures the $K-$way entanglement of subsystem $p$
with its complement.

Versraete et al. \cite{vers02}. have shown that a pure state of four qubits
can be transformed by Stochastic local operations and classical
communication (SLOCC) to a state belonging to one of a set of nine families
of states. In the next two sections, we focus on the calculation of partial $%
K-$way negativities for states representing the nine families of states. The
analytical and numerical results are a pointer to the similarities and
principle differences between the different sets of four qubit states. In
the case of two parameter states, partial $K-$way negativity contours are
plotted to identify the range of parameter values for which a particular $K-$%
way entanglement mode is dominant. We use the notation of ref. \cite{vers02}
to represent different families of four qubit states without going into the
detailed meaning of the symbols used. The nine families of states are
grouped in to two main classes on the basis of partial $K-$way negativities
associated with each family of states. Class I includes all the states
having zero partial $3-$way negativities. The most general states, having
non zero partial $K$-way negativities for $K=2$, $3$ and $4,$ belong to
class II, which also includes three other states characterized by $E_{3}\neq
0$. The states having genuine $K-$partite entanglement of a single type,
that is four partite, tripartite, or bipartite are obtained for particular
parameter values.

\section{Class I - $E_{3}=0$}

The families of states, G$_{abcd}$, L$_{abc_{2}}$, L$_{a_{2}b_{2}}$, and L$%
_{a_{2}0_{3\oplus 1}}$ of ref. \cite{vers02} belong in class I. A common
feature of these states is a positive three way partial transpose,
independent of the qubit with respect to which partial transpose is
constructed. In the normal form, the states do not have three qubit
correlations. All the states have genuine $4-$partite entanglement, which is
destroyed when a measurement is made on the state of a single qubit. In
addition the states may have four-partite entanglement due to pairwise
entanglement. In this case, the mixed state $\rho ^{ABC}=Tr_{D}\left( 
\widehat{\rho }\right) $ may, in turn, have W-like tripartite entanglement
if two qubit coherences for qubit pairs $AB$, $AC$, and $BC$ are
simultaneously nonzero. The reduced two qubit mixed states have pairwise
entanglement.

Consider a genearal four qubit state $\widehat{\rho }=\left\vert \Psi
\right\rangle \left\langle \Psi \right\vert $ where 
\begin{eqnarray}
\left\vert \Psi \right\rangle &=&\alpha \left\vert 0000\right\rangle +\beta
\left\vert 0011\right\rangle +\chi \left\vert 0101\right\rangle +\delta
\left\vert 0110\right\rangle  \notag \\
&&+A\left\vert 1111\right\rangle +B\left\vert 1100\right\rangle +C\left\vert
1010\right\rangle +D\left\vert 1001\right\rangle ,  \label{general}
\end{eqnarray}%
with qubits 1,2,3 and 4, held by parties A, B, C and D respectively. With
proper choice of coeficients, as listed in Table I, the state $\left\vert
\Psi \right\rangle $ represents all possible states belonging to class I.

\begin{table}[tbp]
\caption{List of coeficients to represent G$_{abcd}$, L$_{abc_{2}}$, L$%
_{a_{2}b_{2}}$, and L$_{a_{2}0_{3\oplus 1}}$ by a general state $\left\vert
\Psi \right\rangle $. The Schmidt coeficients $\protect\mu _{0}$, $\protect%
\mu _{1},$ and squared global negativity $\left( N_{G}^{A}\right) ^{2}$ are
also listed.}
\label{tab1}%
\begin{tabular}{||l||c||c||c||c||c||c||c||c||c||c||c||}
\hline\hline
$\left\vert \Psi \right\rangle $ & $\alpha $ & $\beta $ & $\chi $ & $\delta $
& $A$ & $B$ & $C$ & $D$ & $\mu _{0}$ & $\mu _{1}$ & $\left( N_{G}^{A}\right)
^{2}$ \\ \hline\hline
G$_{abcd}$ & $\frac{a+d}{2}$ & $\frac{a-d}{2}$ & $\frac{b+c}{2}$ & $\frac{b-c%
}{2}$ & $\frac{a+d}{2}$ & $\frac{a-d}{2}$ & $\frac{b+c}{2}$ & $\frac{b-c}{2}$
& $\frac{1}{2}$ & $\frac{1}{2}$ & $1$ \\ \hline\hline
L$_{abc_{2}}$ & $\frac{a+b}{2}$ & $\frac{a-b}{2}$ & $c$ & $d$ & $\frac{a+b}{2%
}$ & $\frac{a-b}{2}$ & $c$ & $0$ & $\frac{1+d^{2}}{2}$ & $\frac{1-d^{2}}{2}$
& $1-d^{4}$ \\ \hline\hline
L$_{a_{2}b_{2}}$ & $a$ & $c$ & $b$ & $c$ & $a$ & $0$ & $b$ & $0$ & $\frac{1}{%
2}+c^{2}$ & $\frac{1}{2}-c^{2}$ & $1-4c^{4}$ \\ \hline\hline
L$_{a_{2}0_{3\oplus 1}}$ & $a$ & $b$ & $b$ & $b$ & $a$ & $0$ & $0$ & $0$ & $%
1-\left\vert a\right\vert ^{2}$ & $\left\vert a\right\vert ^{2}$ & $4\left(
\left\vert a\right\vert ^{2}-\left\vert a\right\vert ^{4}\right) $ \\ 
\hline\hline
\end{tabular}%
\end{table}
By writing the state $\left\vert \Psi \right\rangle $ in Schmidt form for
qubit $A$, the global negativity of partial transpose with respect to qubit $%
A$ is found to be $N_{G}^{A}=2\sqrt{\mu _{0}\mu _{1}},$where%
\begin{equation}
\mu _{0}=\left( \left\vert \alpha \right\vert ^{2}+\left\vert \beta
\right\vert ^{2}+\left\vert \chi \right\vert ^{2}+\left\vert \delta
\right\vert ^{2}\right) ,\qquad \text{and \qquad }\mu _{1}=\left( \left\vert
A\right\vert ^{2}+\left\vert B\right\vert ^{2}+\left\vert C\right\vert
^{2}+\left\vert D\right\vert ^{2}\right) .
\end{equation}%
The Schmidt coeficients $\mu _{0}$, $\mu _{1},$ and squared global
negativity $\left( N_{G}^{A}\right) ^{2}$ are also listed in table \ref{tab1}%
. The eigenvector of partially transposed operator $\widehat{\rho }%
_{G}^{T_{A}}$ corresponding to negative eigen value $\lambda ^{-}=-\sqrt{\mu
_{0}\mu _{1}}$ reads as%
\begin{eqnarray}
\left\vert \Psi _{G}^{A-}\right\rangle  &=&\frac{1}{\sqrt{2}}\left( \frac{%
\alpha }{\sqrt{\mu _{0}}}\left\vert 1000\right\rangle +\frac{\beta }{\sqrt{%
\mu _{0}}}\left\vert 1011\right\rangle +\frac{\chi }{\sqrt{\mu _{0}}}%
\left\vert 1101\right\rangle +\frac{\delta }{\sqrt{\mu _{0}}}\left\vert
1110\right\rangle \right)   \notag \\
&&-\frac{1}{\sqrt{2}}\left( \frac{A}{\sqrt{\mu _{1}}}\left\vert
0111\right\rangle +\frac{B}{\sqrt{\mu _{1}}}\left\vert 0100\right\rangle +%
\frac{C}{\sqrt{\mu _{1}}}\left\vert 0010\right\rangle +\frac{D}{\sqrt{\mu
_{1}}}\left\vert 0001\right\rangle \right) .
\end{eqnarray}%
Next, we construct $\widehat{\rho }_{4}^{T_{A}}$ and $\widehat{\rho }%
_{2}^{T_{A}}$ by applying the constraints given in Eqs. (\ref{ptk1},\ref%
{ptk2}) and (\ref{pt21},\ref{pt22}). It is straight forward to obtain 
\begin{eqnarray}
E_{4}^{A} &=&-2\left\langle \Psi _{G}^{A-}\right\vert \widehat{\rho }%
_{4}^{T_{A}}\left\vert \Psi _{G}^{A-}\right\rangle   \notag \\
&=&\frac{4}{N_{G}^{A}}\left( \left\vert A\right\vert ^{2}\left\vert \alpha
\right\vert ^{2}+\left\vert B\right\vert ^{2}\left\vert \beta \right\vert
^{2}+\left\vert C\right\vert ^{2}\left\vert \chi \right\vert ^{2}+\left\vert
D\right\vert ^{2}\left\vert \delta \right\vert ^{2}\right) ,  \label{e4A}
\end{eqnarray}%
and%
\begin{equation}
E_{2}^{A}=-2\left\langle \Psi _{G}^{A-}\right\vert \widehat{\rho }%
_{2}^{T_{A}}\left\vert \Psi _{G}^{A-}\right\rangle
=N_{G}^{A}-E_{4}^{A},\quad E_{0}^{A}=0,  \label{e2A}
\end{equation}%
giving $\allowbreak $%
\begin{equation}
N_{G}^{A}E_{4}^{A}=4\left( \left\vert A\right\vert ^{2}\left\vert \alpha
\right\vert ^{2}+\left\vert B\right\vert ^{2}\left\vert \beta \right\vert
^{2}+\left\vert C\right\vert ^{2}\left\vert \chi \right\vert ^{2}+\left\vert
D\right\vert ^{2}\delta ^{2}\right) ,
\end{equation}%
\begin{eqnarray}
N_{G}^{A}E_{2}^{A} &=&4\left\vert A\right\vert ^{2}\left( \left\vert \beta
\right\vert ^{2}+\left\vert \delta \right\vert ^{2}+\left\vert \chi
\right\vert ^{2}\right) +4\left\vert B\right\vert ^{2}\left( \left\vert
\delta \right\vert ^{2}+\left\vert \alpha \right\vert ^{2}+\allowbreak
\left\vert \chi \right\vert ^{2}\right)   \notag \\
&&+4\allowbreak \left\vert C\right\vert ^{2}\left( \left\vert \beta
\right\vert ^{2}+\left\vert \alpha \right\vert ^{2}+\left\vert \delta
\right\vert ^{2}\right) +4\left\vert D\right\vert ^{2}\left( \left\vert
\alpha \right\vert ^{2}+\allowbreak \left\vert \beta \right\vert
^{2}+\left\vert \chi \right\vert ^{2}\right) .
\end{eqnarray}

\subsection{Monogamy inequalities for four qubit states}

We notice that for four qubit state $\left\vert \Psi \right\rangle $ of Eq. (%
\ref{general}),%
\begin{equation}
\left( N_{G}^{A}\right) ^{2}=N_{G}^{A}E_{4}^{A}+N_{G}^{A}E_{2}^{A},
\end{equation}%
leading to the monogamy inequalities%
\begin{equation}
N_{G}^{A}E_{4}^{A}\leq \left( N_{G}^{A}\right) ^{2}\text{,\quad }%
N_{G}^{A}E_{2}^{A}\leq \left( N_{G}^{A}\right) ^{2}\text{.}
\end{equation}%
The partial $K-$way negativities calculated for the state $\left\vert \Psi
\right\rangle $ represent the amount of entanglement lost when a measurement
is made on the state of a particular qubit. Consider a projective
measurement of qubit $D$, using measurement operators $M_{0}=\left\vert
0\right\rangle \left\langle 0\right\vert $ and $M_{1}=\left\vert
1\right\rangle \left\langle 1\right\vert $. The resulting three qubit state
is a pure state decomposition ($PSD$) written as 
\begin{equation}
\rho _{PSD}^{ABC}=P_{0}\left\vert \Phi _{0}\right\rangle \left\langle \Phi
_{0}\right\vert +P_{1}\left\vert \Phi _{1}\right\rangle \left\langle \Phi
_{1}\right\vert ,
\end{equation}%
where the qubits $ABC$ are found in state 
\begin{equation}
\left\vert \Phi _{0}\right\rangle =\frac{1}{\sqrt{P_{0}}}\left( \alpha
\left\vert 000\right\rangle +\delta \left\vert 011\right\rangle +B\left\vert
110\right\rangle +C\left\vert 101\right\rangle \right) ,
\end{equation}%
with probability $P_{0}=\left\vert B\right\vert ^{2}+\left\vert C\right\vert
^{2}+\left\vert \alpha \right\vert ^{2}+\left\vert \delta \right\vert ^{2},$
and the three qubit state 
\begin{equation}
\left\vert \Phi _{1}\right\rangle =\frac{1}{\sqrt{P_{1}}}\left( \beta
\left\vert 001\right\rangle +\chi \left\vert 010\right\rangle +A\left\vert
111\right\rangle +D\left\vert 100\right\rangle \right) ,
\end{equation}%
occurs with probability $P_{1}=\left\vert A\right\vert ^{2}+\left\vert
D\right\vert ^{2}+\left\vert \beta \right\vert ^{2}+\left\vert \chi
\right\vert ^{2}$. Entanglement lost on measuring the state of qubit $D$ is $%
N_{G}^{D}E_{4}^{D}+N_{G}^{D}E_{2}^{D}$. Defining the\ global negativity of
state $\rho _{PSD}^{ABC}$ as 
\begin{equation}
\left[ N_{PSDG}^{A}(\rho ^{ABC})\right] ^{2}=\left[ N_{G}^{A}(P_{0}\left%
\vert \Phi _{0}\right\rangle \left\langle \Phi _{0}\right\vert )\right] ^{2}+%
\left[ N_{G}^{A}(P_{1}\left\vert \Phi _{1}\right\rangle \left\langle \Phi
_{1}\right\vert )\right] ^{2},
\end{equation}%
we obtain 
\begin{eqnarray}
\left[ N_{PSDG}^{A}(\rho ^{ABC})\right] ^{2} &=&4\left( B^{2}+C^{2}\right)
\left( \alpha ^{2}+\delta ^{2}\right) +4\left( A^{2}+D^{2}\right) \left(
\beta ^{2}+\chi ^{2}\right)   \notag \\
&=&\left[ N_{G}^{A}\right] ^{2}-N_{G}^{A}E_{4}^{A}-N_{G}^{A}E_{2}^{A-AD} 
\notag \\
&=&N_{G}^{A}E_{2}^{A}-N_{G}^{A}E_{2}^{A-AD}.
\end{eqnarray}

Distinct features of states belonging to families G$_{abcd}$, L$_{abc_{2}}$,
L$_{a_{2}b_{2}}$, and L$_{a_{2}0_{3\oplus 1}}$ are discussed below. 
\begin{figure}[t]
\centering \includegraphics[width=5in,height=5.5in,angle=-90]{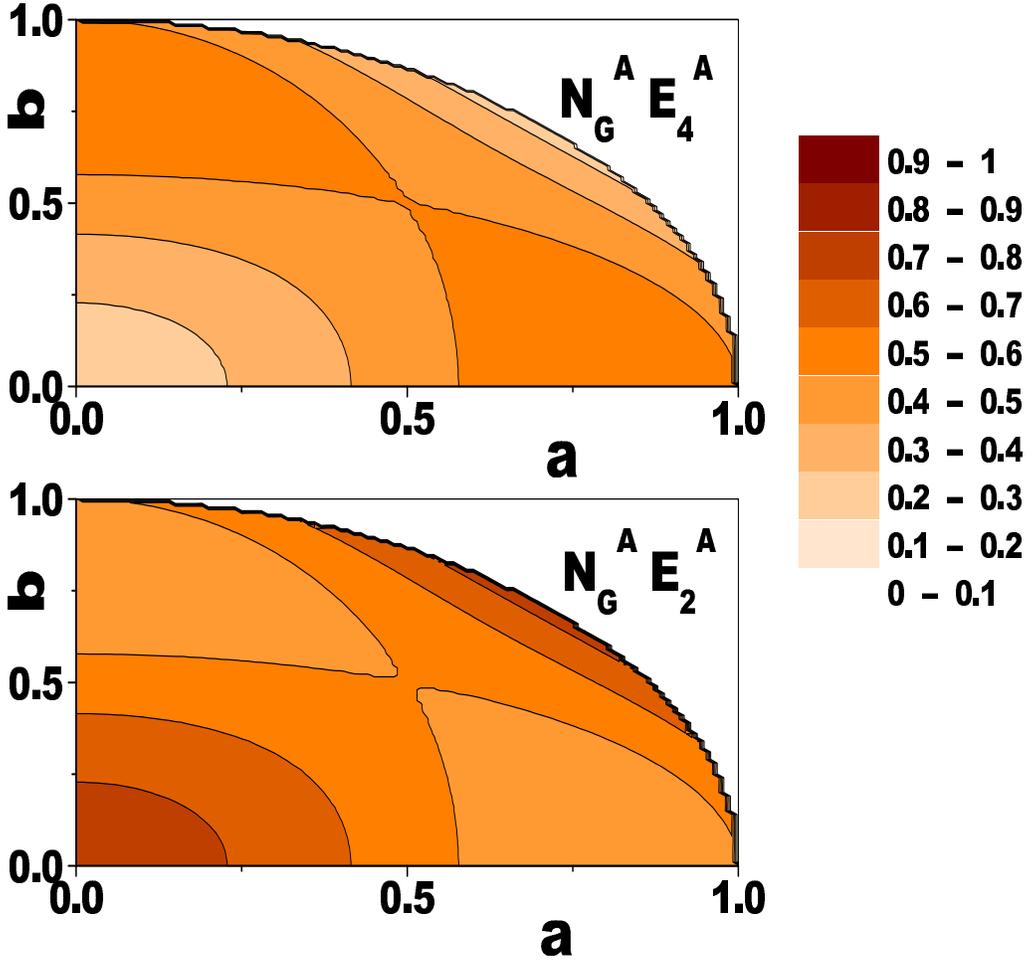}
\caption{Contour plots displaying $N_{G}^{A}E_{4}^{A},$ and $%
N_{G}^{A}E_{2}^{A}$ for the states $G_{abcd}$ as a function of parameters $a$
and $b$.}
\label{fig1}
\end{figure}

\subsection{Set of states G$_{abcd}$}

The set of states G$_{abcd}$%
\begin{eqnarray}
G_{abcd} &=&\frac{a+d}{2}\left( \left\vert 0000\right\rangle +\left\vert
1111\right\rangle \right) +\frac{a-d}{2}\left( \left\vert 1100\right\rangle
+\left\vert 0011\right\rangle \right)  \notag \\
&&+\frac{b+c}{2}\left( \left\vert 1010\right\rangle +\left\vert
0101\right\rangle \right) +\frac{b-c}{2}\left( \left\vert 0110\right\rangle
+\left\vert 1001\right\rangle \right) ,  \label{gabcd}
\end{eqnarray}%
are characterized by $N_{G}^{A}=1$ and $E_{3}^{A}=0$. This result is
consistent with the observation \cite{vers02} that the 3-tangle \cite{coff00}
of the mixed states obtained by tracing out a single qubit of $G_{abcd}$
state is always equal to zero. Substituting $\alpha =A=\frac{a+d}{2}$, $%
\beta =B=\frac{a-d}{2}$,$\chi =C=\frac{b+c}{2}$, and $\delta =D=\frac{b-c}{2}
$ in Eqs. (\ref{e4A}) and (\ref{e2A}) we get%
\begin{equation}
E_{4}^{A}=\frac{1}{4}\left( \left\vert a+d\right\vert ^{4}+\left\vert
a-d\right\vert ^{4}+\left\vert b+c\right\vert ^{4}+\left\vert b-c\right\vert
^{4}\right) ,
\end{equation}%
and $E_{2}^{A}=1-E_{4}^{A}.$ Using the equality%
\begin{equation}
\widehat{\rho }_{2}^{T_{A}}=\widehat{\rho }_{2}^{T_{A-AB}}+\widehat{\rho }%
_{2}^{T_{A-AC}}+\widehat{\rho }_{2}^{T_{A-AD}}-2\widehat{\rho },
\end{equation}%
we further split $E_{2}^{A}$ as 
\begin{eqnarray}
E_{2}^{A} &=&-2\left\langle \Psi _{G}^{A-}\right\vert \widehat{\rho }%
_{2}^{T_{A-AB}}\left\vert \Psi _{G}^{A-}\right\rangle -2\left\langle \Psi
_{G}^{A-}\right\vert \widehat{\rho }_{2}^{T_{A-AC}}\left\vert \Psi
_{G}^{A-}\right\rangle \\
&&-2\left\langle \Psi _{G}^{A-}\right\vert \widehat{\rho }%
_{2}^{T_{A-AD}}\left\vert \Psi _{G}^{A-}\right\rangle +4\left\langle \Psi
_{G}^{A-}\right\vert \widehat{\rho }\left\vert \Psi _{G}^{A-}\right\rangle ,
\end{eqnarray}%
obtaining 
\begin{eqnarray}
E_{2}^{A-AB} &=&\frac{1}{2}\left( \left\vert a+d\right\vert ^{2}\left\vert
a-d\right\vert ^{2}+\left\vert b+c\right\vert ^{2}\left\vert b-c\right\vert
^{2}\right) ,  \notag \\
E_{2}^{A-AC} &=&\frac{1}{2}\left( \left\vert a+d\right\vert ^{2}\left\vert
b+c\right\vert ^{2}+\left\vert a-d\right\vert ^{2}\left\vert b-c\right\vert
^{2}\right) ,  \notag \\
E_{2}^{A-AD} &=&\allowbreak \frac{1}{2}\left( \left\vert a+d\right\vert
^{2}\left\vert b-c\right\vert ^{2}+\left\vert a-d\right\vert ^{2}\left\vert
b+c\right\vert ^{2}\right) .
\end{eqnarray}%
The states have genuine $4-$partite entanglement but no genuine tripartite
entanglement. Fig. (1) displays $N_{G}^{A}E_{4}^{A}$, and $%
N_{G}^{A}E_{2}^{A} $ for the states $G_{abcd}$ for the special case where
coefficients $a$, $b$, and $d=c=\sqrt{\left( 1-a^{2}-b^{2}\right) /2}$ are
real coefficients.

\subsubsection{Entanglement lost on measuring the state of qubit D in state $%
G_{abcd}$}

On tracing over qubit $D$, we get the three qubit state 
\begin{equation}
\rho _{PSD}^{ABC}=Tr_{D}(G_{abcd})=\frac{1}{2}\left[ \left\vert
W_{0}\right\rangle \left\langle W_{0}\right\vert +\left( \left\vert
W_{1}\right\rangle \left\langle W_{1}\right\vert \right) \right] ,
\end{equation}%
which is a mixture of normalized W-like states%
\begin{equation}
\left\vert W_{0}\right\rangle =2\left[ \frac{a+d}{2}\left\vert
000\right\rangle +\frac{a-d}{2}\left\vert 110\right\rangle +\frac{b+c}{2}%
\left\vert 101\right\rangle +\frac{b-c}{2}\left\vert 011\right\rangle ,%
\right]
\end{equation}%
and 
\begin{equation}
\left\vert W_{1}\right\rangle =2\left[ \frac{a+d}{2}\left\vert
111\right\rangle +\frac{a-d}{2}\left\vert 001\right\rangle +\frac{b+c}{2}%
\left\vert 010\right\rangle +\frac{b-c}{2}\left\vert 100\right\rangle \right]
.
\end{equation}%
The loss of qubit $D$ results in total loss of four-partite entanglement of
the state\ $G_{abcd}$. Besides that the two-way coherences involving qubit $%
D $ are also annihilated. Recalling that in the state $G_{abcd}$ all qubits
have equal amount of $K-$way coherences, the mixed state $\rho _{PSD}^{ABC}$
has%
\begin{equation}
\left[ N^{A}(\rho ^{ABC})\right] ^{2}\leq \left(
N_{G}^{A}E_{2}^{A-AB}+N_{G}^{A}E_{2}^{A-AC}\right)
\end{equation}%
where 
\begin{equation}
N_{G}^{A}E_{2}^{A-AB}+N_{G}^{A}E_{2}^{A-AC}=\left( \left\vert a\right\vert
^{2}+\left\vert d\right\vert ^{2}\right) \left( \left\vert b\right\vert
^{2}+\left\vert c\right\vert ^{2}\right)
\end{equation}%
On the other hand if party D measures the state of qubit $D$ and
communicates classically to parties $A$, $B$, and $C$, W-like entangled
states become available to parties $A$, $B$, and $C$. Qubit $D$ is found in
state $\left\vert 0\right\rangle $, with probability $\frac{1}{2}$ leaving
the qubits $ABC$ in normalized state $\left\vert W_{0}\right\rangle ,$
whereas the result $\left\vert 1\right\rangle $ for the state of qubit $D$
collapses the four qubit state to three qubit state $\left\vert
W_{1}\right\rangle $. Defining $\rho _{W_{0}}=\left\vert W_{0}\right\rangle
\left\langle W_{0}\right\vert $ and $\rho _{W_{1}}=\left\vert
W_{1}\right\rangle \left\langle W_{1}\right\vert $, it is found that 
\begin{equation}
\left[ N_{G}^{B}\left( \frac{1}{2}\rho _{W_{0}}\right) \right] ^{2}+\left[
N_{G}^{B}\left( \frac{1}{2}\rho _{W_{1}}\right) \right] ^{2}=\left(
\left\vert a\right\vert ^{2}+\left\vert d\right\vert ^{2}\right) \left(
\left\vert b\right\vert ^{2}+\left\vert c\right\vert ^{2}\right) ,
\end{equation}%
as such 
\begin{equation}
\left[ N_{PSDG}^{B}(\rho ^{BCD})\right]
^{2}=N_{G}^{A}E_{2}^{A-AB}+N_{G}^{A}E_{2}^{A-AC}.
\end{equation}%
Entanglement loss due to loss of qubit $D$ is caused by loss of information
about the state of qubit $D$. On further state reduction, two qubit
entangled states are obtained from the states $\left\vert W_{0}\right\rangle 
$, and $\left\vert W_{1}\right\rangle $.

\subsection{Set of states L$_{abc_{2}}$}

For the set of normalized states%
\begin{eqnarray}
L_{abc_{2}} &=&\frac{a+b}{2}\left( \left\vert 0000\right\rangle +\left\vert
1111\right\rangle \right) +\frac{a-b}{2}\left( \left\vert 1100\right\rangle
+\left\vert 0011\right\rangle \right)   \notag \\
&&+c\left( \left\vert 1010\right\rangle +\left\vert 0101\right\rangle
\right) +d\left\vert 0110\right\rangle ,  \label{TWO}
\end{eqnarray}%
with $d^{2}=1-2\left\vert c\right\vert ^{2}-\left\vert b\right\vert
^{2}-\left\vert a\right\vert ^{2}$ we get $\left( N_{G}^{A}\right)
^{2}=1-d^{4}$. The product of global negativity and $4-$way negativity is
found to be 
\begin{equation}
N_{G}^{A}E_{4}^{A}=4\left( \left\vert \frac{a+b}{2}\right\vert
^{4}+\left\vert \frac{a-b}{2}\right\vert ^{4}+\left\vert c\right\vert
^{4}\right) ,
\end{equation}%
where as the product 
\begin{equation}
N_{G}^{A}E_{2}^{A}=N_{G}^{A}\left( E_{2}^{AB}+E_{2}^{AC}+E_{2}^{AD}\right) ,
\end{equation}%
with 
\begin{equation}
N_{G}^{A}E_{2}^{AB}=\frac{1}{2}\left\vert a+b\right\vert ^{2}\left\vert
a-b\right\vert ^{2}+\left\vert c\right\vert ^{2}d^{2},
\end{equation}%
\begin{equation}
N_{G}^{A}E_{2}^{AC}=2\left\vert a+b\right\vert ^{2}\left\vert c\right\vert
^{2}+\left\vert a-b\right\vert ^{2}d^{2},
\end{equation}%
and%
\begin{equation}
N_{G}^{A}E_{2}^{AD}=2\left\vert a-b\right\vert ^{2}\left\vert c\right\vert
^{2}+\left\vert a+b\right\vert ^{2}d^{2}.
\end{equation}

\subsection{Sets of states $L_{a_{2}b_{2}}$ and L$_{a_{2}0_{3\oplus 1}}$}

The general form for the family of states $L_{a_{2}b_{2}}$ is%
\begin{equation}
L_{a_{2}b_{2}}=a\left( \left\vert 0000\right\rangle +\left\vert
1111\right\rangle \right) +b\left( \left\vert 0101\right\rangle +\left\vert
1010\right\rangle \right) +c\left( \left\vert 0110\right\rangle +\left\vert
0011\right\rangle \right) ,  \label{THREE}
\end{equation}%
where%
\begin{equation*}
c^{2}=\frac{1-2\left\vert a\right\vert ^{2}-2\left\vert b\right\vert ^{2}}{2}%
,\quad 0\leq \left\vert b\right\vert \leq \frac{1}{\sqrt{2}},\quad 0\leq
\left\vert a\right\vert \leq \frac{1}{\sqrt{2}}.
\end{equation*}%
The calculated squared Global negativity is given by $\left(
N_{G}^{A}\right) ^{2}=1-4c^{4}$ and $N_{G}^{A}E_{4}^{A}=4\left( \left\vert
a\right\vert ^{4}+\left\vert b\right\vert ^{4}\right) $. The pairwise
partial negativities read as%
\begin{equation}
N_{G}^{A}E_{2}^{AB}=N_{G}^{A}E_{2}^{AD}=4c^{2}\left( \left\vert a\right\vert
^{2}+\left\vert b\right\vert ^{2}\right) ,\quad
N_{G}^{A}E_{2}^{AC}=8\left\vert a\right\vert ^{2}\left\vert b\right\vert
^{2}.
\end{equation}

The states%
\begin{equation}
L_{a_{2}0_{3\oplus 1}}=a\left( \left\vert 0000\right\rangle +\left\vert
1111\right\rangle \right) +b\left( \left\vert 0101\right\rangle +\left\vert
0110\right\rangle +\left\vert 0011\right\rangle \right)  \label{FOUR}
\end{equation}%
have%
\begin{equation}
\left( N_{G}^{A}\right) ^{2}=4\left( a^{4}+3a^{2}b^{2}\right) ,\quad
N_{G}^{A}E_{4}^{A}=4a^{4},\quad N_{G}^{A}E_{2}^{A}=12\left\vert a\right\vert
^{2}\left\vert b\right\vert ^{2},
\end{equation}%
with 
\begin{equation}
N_{G}^{A}E_{2}^{AB}=N_{G}^{A}E_{2}^{AC}=N_{G}^{A}E_{2}^{AD}=4\left\vert
a\right\vert ^{2}\left\vert b\right\vert ^{2}.
\end{equation}%
and $tr_{D}\left( \left\vert L_{a_{2}0_{3\oplus 1}}\right\rangle
\left\langle L_{a_{2}0_{3\oplus 1}}\right\vert \right) $ desplays W-like
entanglement. That means we can extract pairwise entanglement from the
normal form.

\section{Class II - $E_{3}\neq 0$}

The sets of states L$_{ab_{3}}$ and L$_{a_{4}}$, having 4-partite,
tripartite and bi-partite entanglement are grouped in class II, along with
the states L$_{0_{5\oplus 3}}$, L$_{0_{7\oplus 1}}$, and L$_{0_{3\oplus
1}0_{3\oplus 1}}$ having $E_{3}\neq 0$ while $E_{4}=0.$

\begin{figure}[t]
\centering \includegraphics[width=5in,height=5.5in,angle=-90]{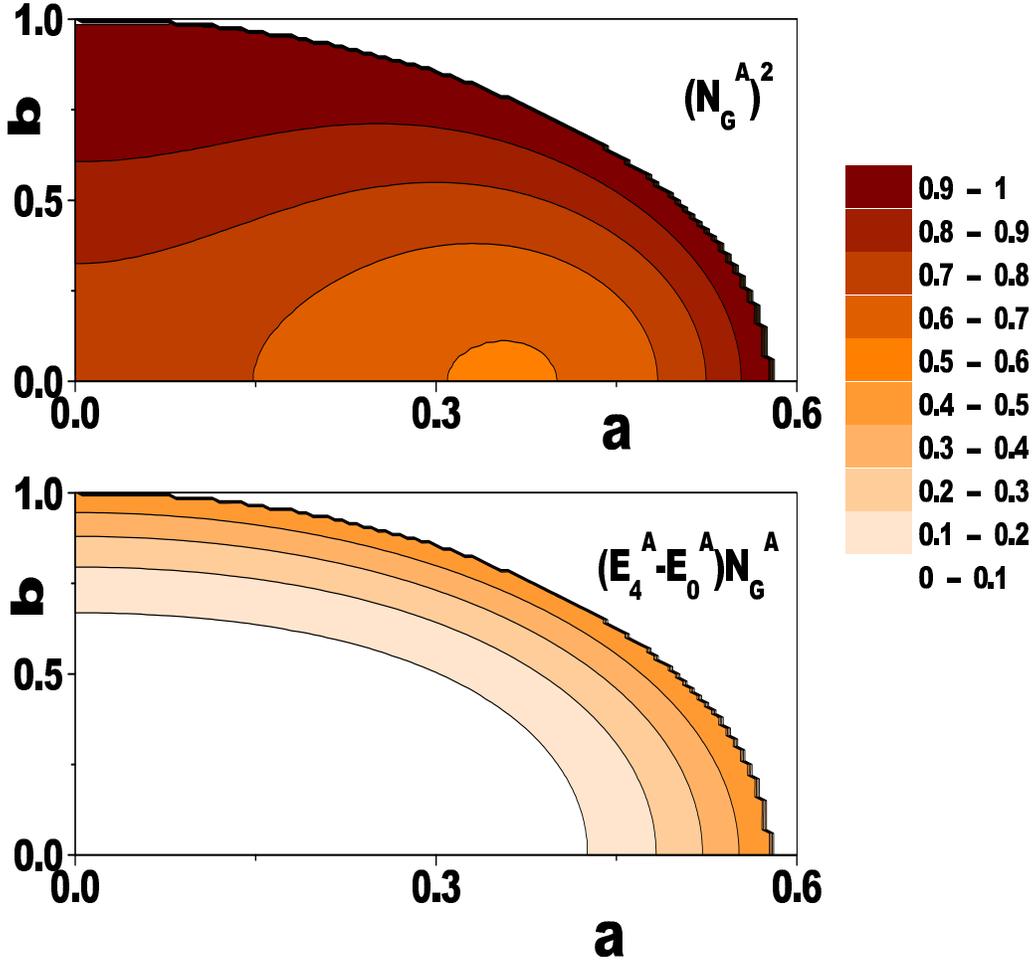}
\caption{Contour plots displaying $N_{G}^{A}(E_{4}^{A}-E_{0}^{A}),$ and $%
(N_{G}^{A})^{2}$ for the states $L_{ab3}$ as a function of parameters $a$
and $b$.}
\label{fig2}
\end{figure}

\begin{figure}[t]
\centering \includegraphics[width=5in,height=5.5in,angle=-90]{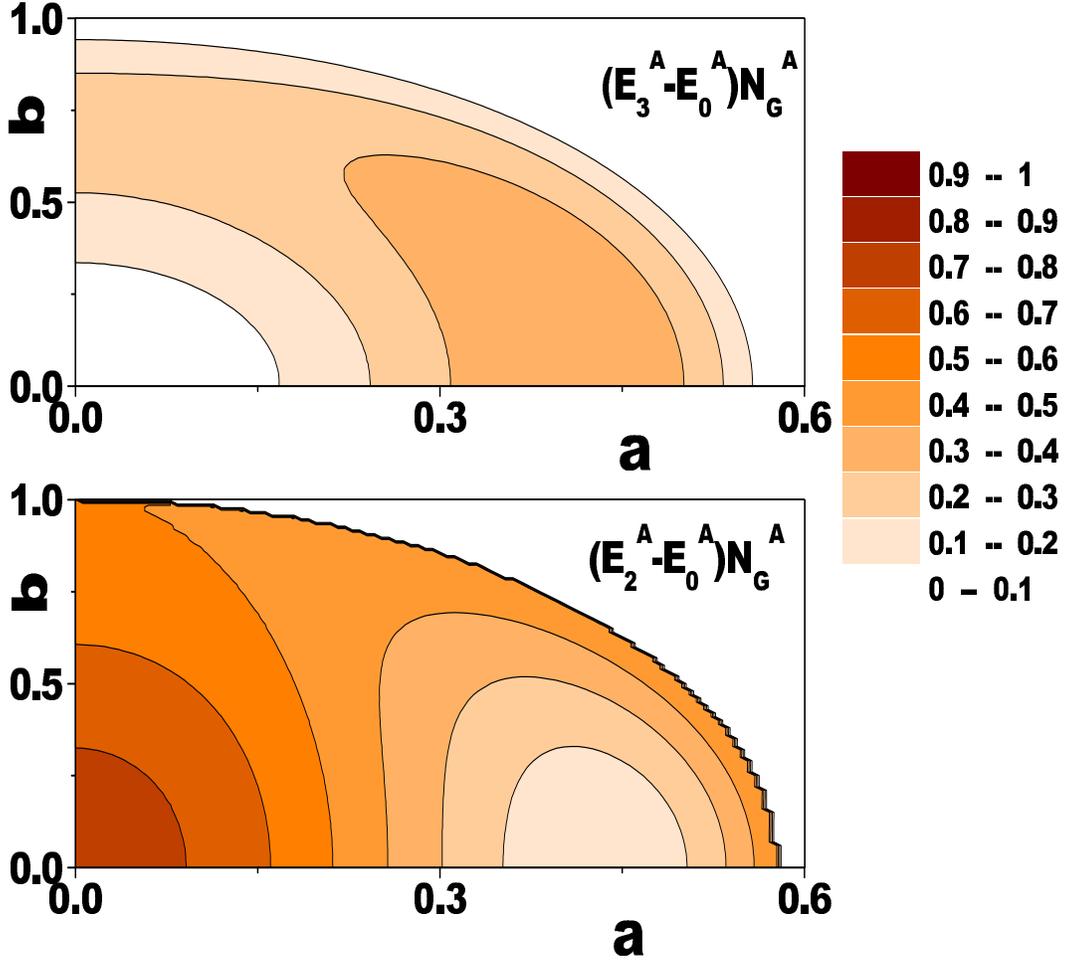}
\caption{Contour plots displaying $N_{G}^{A}(E_{3}^{A}-E_{0}^{A})$ and $%
N_{G}^{A}(E_{2}^{A}-E_{0}^{A})$ for the states $L_{ab_{3}}$ as a function of
parameters $a$ and $b$. }
\label{fig3}
\end{figure}

\subsection{Sets of states L$_{ab_{3}}$ and L$_{a_{4}}$}

The states in the family L$_{ab_{3}}$have bipartite, tripartite as well as
4-partite entanglement. The normalized two parameter state has the general
form%
\begin{eqnarray}
L_{ab_{3}} &=&a\left( \left\vert 0000\right\rangle +\left\vert
1111\right\rangle \right) +\frac{a+b}{2}\left( \left\vert 0101\right\rangle
+\left\vert 1010\right\rangle \right)  \notag \\
&&+\frac{a-b}{2}\left( \left\vert 0110\right\rangle +\left\vert
1001\right\rangle \right)  \notag \\
&&+\frac{ic}{\sqrt{2}}\left( \left\vert 0001\right\rangle +\left\vert
0010\right\rangle +\left\vert 0111\right\rangle +\left\vert
1011\right\rangle \right) ,  \label{lab3_FIVE}
\end{eqnarray}%
where $0\leq \left\vert a\right\vert \leq \frac{1}{\sqrt{3}},$ $0\leq
\left\vert b\right\vert \leq 1$, and $c=\sqrt{\left( 1-b^{2}-3a^{2}\right)
/2.0}$ . Figs. (\ref{fig2}) and (\ref{fig3}) display the contour plots of
global negativity and partial two, three and four way negativities
calculated numerically from state operator partially transposed with respect
to qubit $A$. For the two parameter states $L_{ab_{3}}$, maximum value of $%
N_{G}^{A}E_{4}^{A}=0.5$, occurs for $a=\frac{1}{\sqrt{3}},$ $b=0,$ $c=0$
that is for the state 
\begin{eqnarray}
\Psi &=&\frac{1}{\sqrt{3}}\left( \left\vert 0000\right\rangle +\left\vert
1111\right\rangle \right) +\frac{1}{2\sqrt{3}}\left( \left\vert
0101\right\rangle +\left\vert 1010\right\rangle \right)  \notag \\
&&+\frac{1}{2\sqrt{3}}\left( \left\vert 0110\right\rangle +\left\vert
1001\right\rangle \right) \text{.}
\end{eqnarray}

\begin{figure}[t]
\centering \includegraphics[width=3.75in,height=5in,angle=-90]{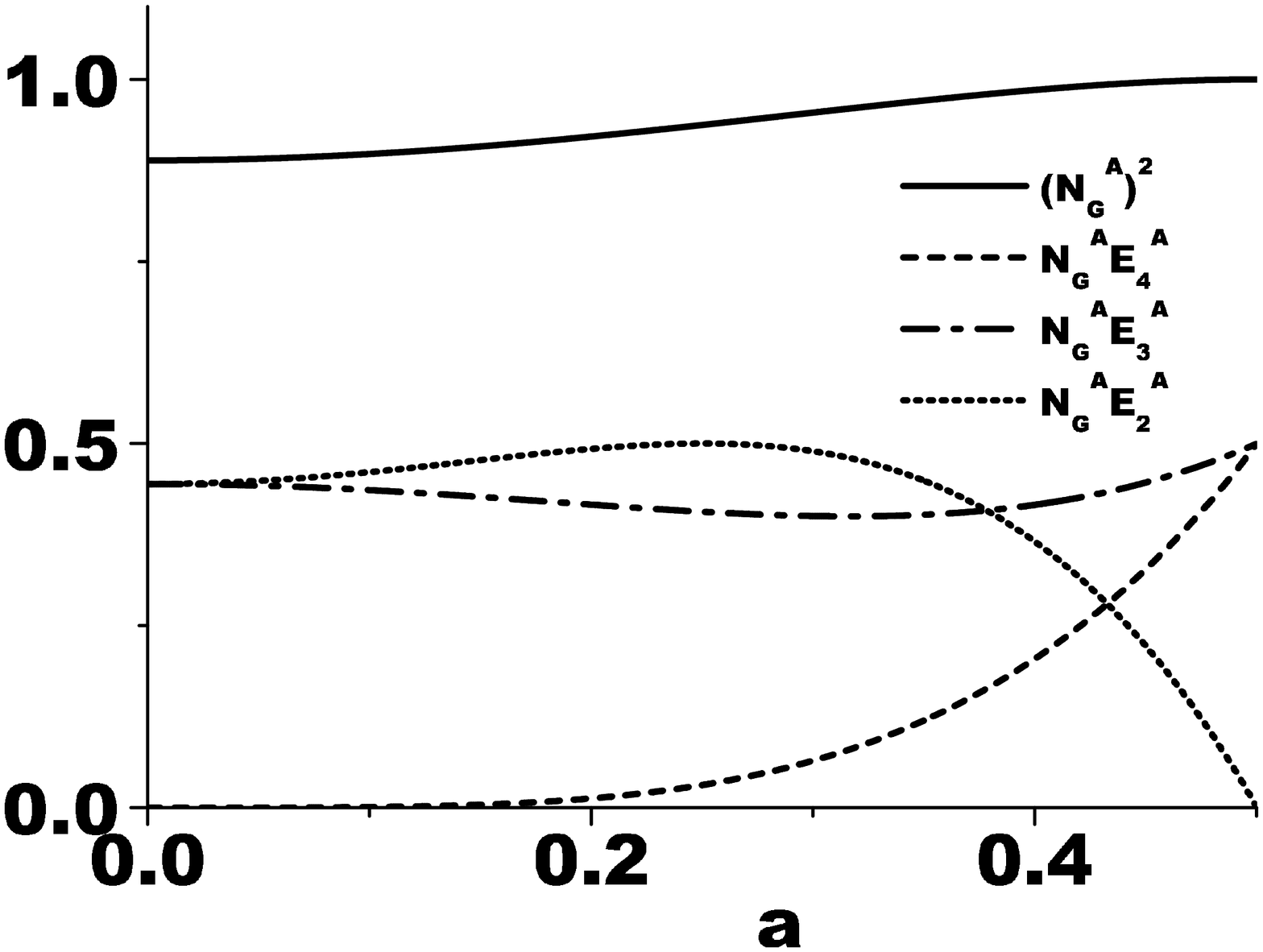}
\caption{Global and partial negativities for qubit $A$ versus parameter $a$
for the states L$_{a_{4}}$.}
\label{fig4}
\end{figure}
\begin{figure}[t]
\centering \includegraphics[width=3.75in,height=5in,angle=-90]{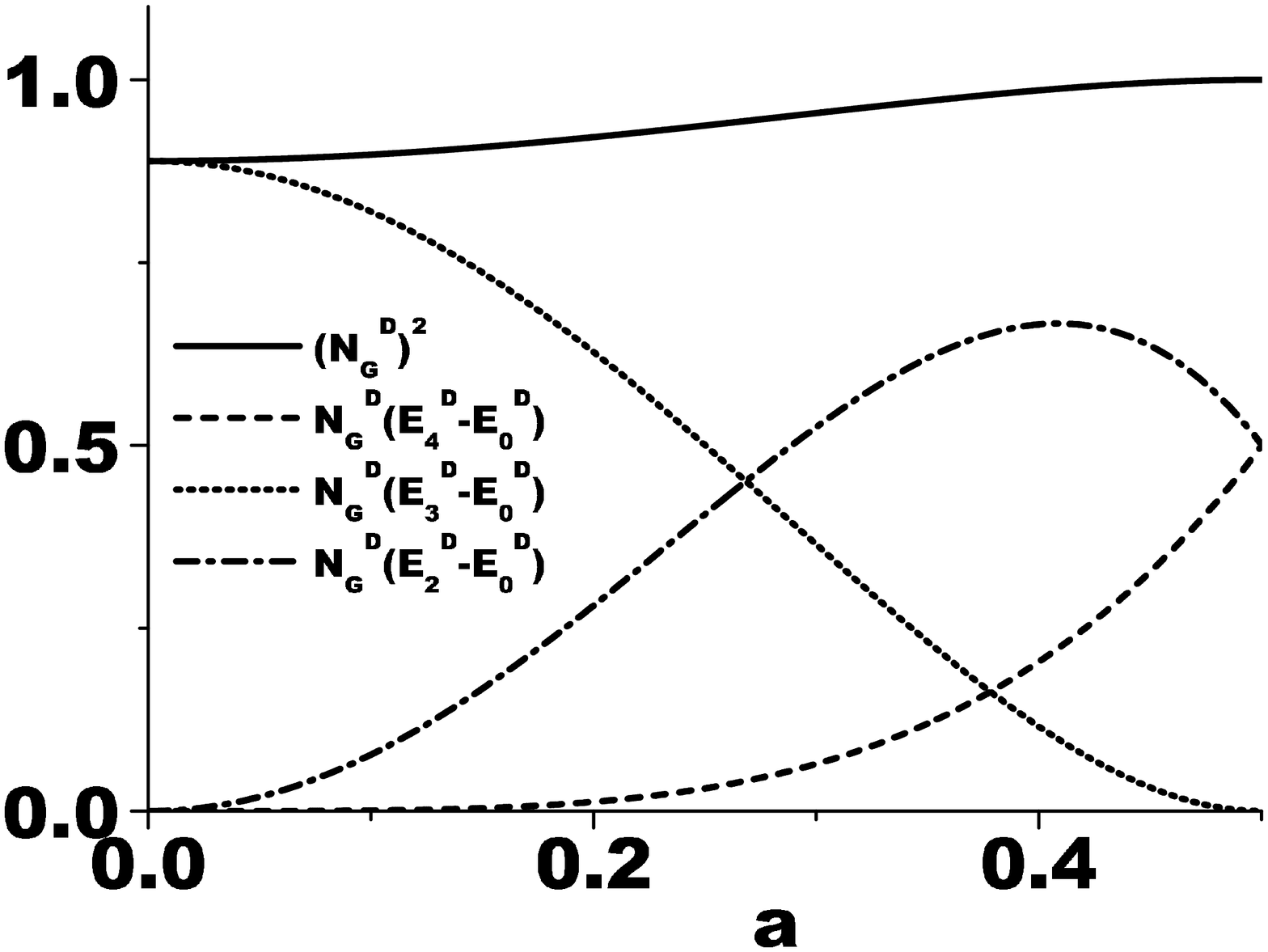}
\caption{Global and partial negativities for qubit $D$ versus  parameter $a$
for the states L$_{a_{4}}$.}
\label{fig5}
\end{figure}

Single parameter states

\begin{eqnarray}
L_{a_{4}} &=&a\left( \left\vert 0000\right\rangle +\left\vert
0101\right\rangle +\left\vert 1010\right\rangle +\left\vert
1111\right\rangle \right)  \notag \\
&&\sqrt{\frac{1-4a^{2}}{3}}\left( i\left\vert 0001\right\rangle +\left\vert
0110\right\rangle -i\left\vert 1011\right\rangle \right) ,  \label{SIX}
\end{eqnarray}%
are also characterized by partial two, three and four way negativities such
that 
\begin{equation}
\left( N_{G}^{A}\right) ^{2}=\frac{8}{9}\left( a^{2}-2a^{4}+1\right) ,\quad
N_{G}^{A}E_{4}^{A}=8a^{4},
\end{equation}

\begin{equation}
\quad N_{G}^{A}E_{3}^{A}=\frac{4}{9}\left( 4a^{2}-32a^{4}+1\right) ,\quad
N_{G}^{A}E_{2}^{A}=\frac{4}{9}\left( 10a^{4}-2a^{2}+1\right) .
\end{equation}%
The global negativity and partial negativities as a function of parameter $%
0\leq a\leq \frac{1}{2}$ are displayed in Fig. (\ref{fig4}) for qubit $A$
and in Fig. (\ref{fig5}) for qubit $D$. We notice that the state is in
Schmidt like form for qubit $A$ with $E_{0}^{A}=0.$

\subsection{The states L$_{0_{5\oplus 3}}$, L$_{0_{7\oplus 1}}$ and L$%
_{0_{3\oplus 1}0_{3\oplus 1}}$}

A common feature of the states L$_{0_{5\oplus 3}}$, L$_{0_{7\oplus 1}}$ and L%
$_{0_{3\oplus 1}0_{3\oplus 1}}$ is $E_{4}=0$. The state

\begin{equation}
L_{0_{5\oplus 3}}=\frac{1}{2}\left( \left\vert 0000\right\rangle +\left\vert
0101\right\rangle +\left\vert 1000\right\rangle +\left\vert
1110\right\rangle \right)  \label{SEVEN}
\end{equation}%
has only genuine tripartite and bi-partite entanglement, while $\left(
N_{G}^{A}\right) ^{2}=\left( N_{G}^{D}\right) ^{2}=0.75$. The tripartite
entanglement for qubits $ABC$, $ABD$, and $ACD$\ is found to be equal that is%
\begin{equation}
N_{G}^{A}E_{3}^{A-ABC}=N_{G}^{A}E^{A-ABD}=N_{G}^{D}E_{3}^{D-ACD}=0.25,
\end{equation}%
while $N_{G}^{D}E_{2}^{D-BD}=0.25.$ For the mixed state 
\begin{equation}
\rho _{PSD}^{ABC}=Tr_{D}(\left\vert L_{0_{5\oplus 3}}\right\rangle
\left\langle L_{0_{5\oplus 3}}\right\vert )=\frac{3}{4}\left\vert
T_{0}\right\rangle \left\langle T_{0}\right\vert +\frac{1}{4}\left(
\left\vert 010\right\rangle \left\langle 010\right\vert \right) ,
\end{equation}%
where%
\begin{equation*}
T_{0}=\frac{1}{\sqrt{3}}\left( \left\vert 000\right\rangle +\left\vert
100\right\rangle +\left\vert 111\right\rangle \right) ,
\end{equation*}%
the relation%
\begin{equation}
\left[ N_{G}^{A}\left( \frac{3}{4}\left\vert T_{0}\right\rangle \left\langle
T_{0}\right\vert \right) \right] ^{2}+\left[ N_{G}^{A}\left( \frac{1}{4}%
\left( \left\vert 010\right\rangle \left\langle 010\right\vert \right)
\right) \right] ^{2}=N_{G}^{A}E_{3}^{A-ABC},
\end{equation}%
holds. Three tangle of the mixed state is also found to be $0.25$ if qubit $%
B $, $C$ or $D$ is traced out and zero if qubit $A$ is traced out.

Qubit $A$ has genuine tripartite entanglement in state

\begin{equation}
L_{0_{7\oplus 1}}=\frac{1}{2}\left( \left\vert 0000\right\rangle +\left\vert
1011\right\rangle +\left\vert 1101\right\rangle +\left\vert
1110\right\rangle \right) ,  \label{EIGHT}
\end{equation}%
$\allowbreak \allowbreak $giving $\left( N_{G}^{A}\right)
^{2}=N_{G}^{A}E_{3}^{A}=0.75$, and $%
N_{G}^{A}E_{3}^{A-ABC}=N_{G}^{A}E_{3}^{A-ABD}=N_{G}^{A}E_{3}^{A-ACD}=0.25$.
The partial negativities for qubit $D$ are 
\begin{equation}
\left( N_{G}^{D}\right) ^{2}=1.0,\quad N_{G}^{D}E_{3}^{D}=0.5,\quad
N_{G}^{D}E_{2}^{D}=0.5,
\end{equation}%
\begin{equation}
N_{G}^{D}E_{3}^{D-ABD}=N_{G}^{D}E_{3}^{D-ACD}=0.25,\quad
N_{G}^{D}E_{2}^{D-BD}=N_{G}^{D}E_{2}^{D-CD}=0.25.
\end{equation}%
The state $Tr_{D}(\left\vert L_{0_{5\oplus 3}}\right\rangle \left\langle
L_{0_{5\oplus 3}}\right\vert )$, has GHZ like correlations between the
qubits $ABC$, whereas, $Tr_{A}(\left\vert L_{0_{5\oplus 3}}\right\rangle
\left\langle L_{0_{5\oplus 3}}\right\vert )$, has pair wise residual
entanglement.

The product of separable qubit $A$ and three qubit GHZ state constitutes the
state%
\begin{equation}
L_{0_{3\oplus 1}0_{3\oplus 1}}=\frac{1}{\sqrt{2}}\left( \left\vert
0000\right\rangle +\left\vert 0111\right\rangle \right) ,  \label{NINE}
\end{equation}%
with $\left( N_{G}^{A}\right) ^{2}=0$, $\left( N_{G}^{p}\right)
^{2}=N_{G}^{p}E_{3}^{p}=1.0$ for $p=B$ or $C$ or $D$.

\section{Conclusions}

To summarise, the nine families of four qubit states obtained by Versraete
et al. \cite{vers02} can be grouped in two distinct classes. Four sets of
states, G$_{abcd}$ , L$_{abc_{2}},$ L$_{a_{2}b_{2}}$, and L$%
_{a_{2}0_{3\oplus 1}}$, have non zero partial $4-$way and $2-$way
negativities, while $3-$way partial negativities are zero in the normal
form. The families of states L$_{ab_{3}}$ and L$_{a_{4}}$ are distinctly
different from the sets of states in the first category in that the states
have bi, tri as well as 4-partite entanglement in the normal form. The
states L$_{0_{7\oplus 1}}$\ and L$_{0_{5\oplus 3}}$ characterized by $%
E_{4}=0 $, $E_{3}\neq 0$, $E_{2}\neq 0$, along with the state L$_{0_{3\oplus
1}0_{3\oplus 1}}$, having $E_{4}=0$, $E_{3}\neq 0$, $E_{2}=0$, are also
included in Class II. Partial four way negativity is a measure of genuine $%
4- $partite entanglement of the state. Three-way partial negativity
determines the probabilistic entanglement that becomes available to the
three parties after the fourth party measures the state of the qubit it
holds. Two-way negativity measures the pairwise entanglement.

The coefficients in a normal form being local invariants, the partial $k-$%
way negativities are proper entanglement measures satisfying the conditions
of normalization, convexity and monotonicity. Whereas, the global negativity
with respect to a given qubit $p$ gives information about the amount of
multipartite entanglement that is lost on the loss of qubit $p$, the partial 
$K-$way negativities give detailed information about the nature and
distribution of quantum correlations lost due to the loss of a qubit. The
partial $K-$way negativities are meaningful polynomials of local invariants.
Local unitary rotations on the state in normal form may enhance a given set
of partial $K-$way negativities at the cost of others. For a state that is
not in normal form, the partial $K-$way negativities measure the coherences
present in the composite system. The monogamy relations obtained, naturally,
for the 4-partite, tripartite and bipartite entanglement of a given qubit,
provide further insight into entanglement distribution in four qubit states.
We believe that quantifing the multipartite quantum correlations through
partial $K-$way negativities will facilitate the construction and
implementation of quantum information processing protocols.


\begin{thebibliography}{99}
\bibitem{zycz98} K. Zyczkowski, P. Horodecki, A. Sanpera, and M. Lewenstein,
Phys. Rev. A 58, 883 (1998).

\bibitem{pere96} A. Peres, Phys. Rev. Lett. 77, 1413 (1996).

\bibitem{horo96} M. Horodecki, P. Horodecki, and R. Horodecki, Phys. Lett. A
223, 8 (1996).

\bibitem{vida02} G. Vidal and R. F. Werner, Phys. Rev. Vol. 65, 032314
(2002).

\bibitem{eise01} J. Eisert, PhD thesis (University of Potsdam, February
2001).

\bibitem{shar08} S. S. Sharma and N. K. Sharma, Phys. Rev. A 77, 042117
(2008).

\bibitem{shar082} S. S. Sharma and N. K. Sharma, Phys. Rev. A 78, 012113
(2008).

\bibitem{shel06} S. Shelly Sharma and N. K. Sharma, arXiv: quant-ph/0608062
(unpublished).

\bibitem{shar07} S. Shelly Sharma and N. K. Sharma, Phys. Rev. A 76, 012326
(2007).

\bibitem{vers02} F. Versraete, J. Dehaene, B. De Moor, and H. Verschelde,
Phys. Rev. A65, 052112 (2002).

\bibitem{cart99} H. A. Carteret, N. Linden , S. Popescu, and A. Sudbery,
Foundations of Physics, Vol. 29, No. 4, 527 (1999); H. A. Carteret, A.
Higuchi, A. Sudbery, J. Math. Phys. 41, 7932 (2000).

\bibitem{coff00} V. Coffman, J. Kundu, and W. K. Wootters, Phys. Rev. A 61,
052306 (2000).
\end{thebibliography}
\end{document}